# A New Empirical Constraint on the Prevalence of Technological Species in the Universe

- A. Frank[1]  &  W.T. Sullivan, III[2]

*Abstract: In this paper we address the cosmic frequency of technological species. Recent advances in exoplanet studies provide strong constraints on all astrophysical terms in the Drake Equation.   Using these and modifying the form and intent of the Drake equation we show that we can set a firm lower bound on the probability that one or more additional technological species have evolved anywhere and at any time in the history of the observable Universe.  We find that as long as the probability that a habitable zone planet develops a technological species is larger than ~$10^{-24}$, then humanity is not the only time technological intelligence has evolved.  This constraint has important scientific and philosophical consequences.*

**Introduction:** The history of physics has shown that fundamental insights into a problem can sometimes be acquired by setting order of magnitude estimates of scales, limits and boundaries via combinations of constants or parameters.  This approach can be particularly effective for problems in which few empirically derived constraints are available, such as the derivation of the quantum gravitational Planck Length i.e., $l_p = (hG/c^3)^{1/2}$ or early studies using the cosmic density parameter ($\Omega=\rho/\rho_c$).  Here we adopt such an approach to address  an overarching goal of astrobiology: understanding if humanity is "alone." (Impey et al 2012, Sullivan et al 2007) One way of framing this question is to ask if other technology-building species exist now, or have ever existed at anytime, in the Universe. Given that we have only one known example of a planet where such evolution has occurred (and also only one example of where even microbial life has evolved), there seems to be little hope in determining the fecundity of the Universe in producing technological species. This conclusion, however, ignores the rapid and substantial progress astrobiology has made in the last two decades. In particular, the empirical determination of exoplanet statistics has radically changed the nature and quality of constraints astrobiologists now have at their disposal when considering the prevalence of life in the Universe. In this paper, we employ these new constraints to set a lower limit on the probability that technological species have ever evolved anywhere other than on Earth.

[1] Department of Physics & Astronomy, University of Rochester, Rochester, New York, USA
[2] Department of Astronomy & Astrobiology Program, University of Washington, Seattle, Washington, USA

**Method.** Our approach asks a very different question from the usual treatment of the subject. Standard astrobiological discussions of intelligent life focus on how many technological species *currently* exist with which we might communicate (Vakoch et al 2015). But rather than asking whether we are *now* alone, we ask whether we are the only technological species that has *ever* arisen. Such an approach allows us to set limits on what might be called the "cosmic archaeological question": How often in the history of the Universe has evolution ever led to a technological species, whether short- or long-lived? As we shall show, providing constraints on an answer to this question has profound philosophical and practical implications.

We first modify the Drake equation in order to address how many technological species have formed over the history of the observable Universe. We call this number $A$ (for archaeology) and use it to investigate the probability that humanity is unique (i.e., $A = 1$). Note that we are explicitly *not* concerned with the average lifetime *<L>* of such species, or if they still exist such that we could receive their signals or signal them. This is in contrast to the usual Drake equation formulation [3], which calculates $N_c$, the number of technological species *now* existing, (hence its concern with *<L>)*. Given this approach, effects of time are removed where they normally appear in the form of star formation rates, stellar lifetimes, technological species lifetimes, and distribution of epochs of arising. As usual for the Drake equation, we assume that technology is associated with planets and their host stars.

We define the "A-form" of the Drake equation, which describes the total number of technological species that have *ever* evolved anywhere in the currently observable Universe:

$$A = [N_* f_p n_p][f_l f_i f_t]$$

$$= N_{ast} f_{bt} \qquad (1)$$

where $N_*$ is the total number of stars, $f_p$ is the fraction of those stars that form planets, $n_p$ is the average number of planets in the habitable zone of a star with planets, $f_l$ is the probability that a habitable zone planet develops life, $f_i$ is the probability that a planet with life develops intelligence, and $f_t$ is the probability that a planet with intelligent life develops technology (of the "energy intensive" kind such as our own civilization).

The second version of equation (1) reduces the right-hand side to two factors, the first of which, $N_{ast}$, includes all factors involving astrophysics and represents the total number of habitable zone planets. The second factor, $f_{bt}$, gathers the three factors involved with biology, evolution and "planetary sociology," and represents the total "bio-technical" probability that a given habitable zone planet has ever evolved a technological species. The factor $f_{bt} = f_l f_i f_t$ is extremely uncertain, basically because (a) we have no theory to guide any estimates, and (b) we have only one known example of the occurrence and history of life, intelligence and technology. We leave $f_{bt}$ as simply statistically unknown at this time and examine the consequences of it taking on various values depending on one's pessimism or optimism.

In contrast, the key development that enables our new approach is that observations now provide accepted or statistical well-determined values for all factors contributing to $N_{ast}$. The importance of this accomplishment cannot be overstated - until recently only one factor ($N_*$) was known and it was entirely possible that habitable zone planets might have been extremely rare ($N_{ast}/N_* \ll 1$).

But the combination of radial velocity, transit and microlensing based methods now yields statistically well constrained values for both $f_p$ and $n_p$.[3]

We adopt the values $f_p \sim 1.0$ (Cassan et al 2013) and $n_p \sim 0.2$ (Petigura et al 2013), therefore $N_{ast}/N_* = 0.2$. Note that the number of stars $N_*$ is also an observable quantity, which depends on the size of the region being considered.

**Results:** We now turn to the specific question: "Has even one other technological species ever existed in the observable Universe?" We take $N_* = 2 \times 10^{22}$ for the total number of stars in the Universe (Silburt et al 2015) To address our question, $A$ is set to a conservative value ensuring that Earth is the only location in the history of the cosmos where a technological civilization has ever evolved. Adopting $A = 0.01$ means that in a statistical sense were we to rerun the history of the Universe 100 times, only once would a lone technological species occur. A lower bound $\bar{f}_{bt}$ on the probability is then

$$\bar{f}_{bt} = A / N_{ast} = 0.01 / 4 \times 10^{21} = 2.5 \times 10^{-24}$$

Thus only for values of the product $f_{bt}$ lower than $2.5 \times 10^{-24}$ are we likely to be alone and singular in the history of the observable Universe. This limiting value of $f_{bt}$ can be considered to define a "pessimism line" in discussions of the prevalence of technological civilizations on a cosmic scale. On the other hand if evolutionary processes lead to higher values, then we can be assured that we are not the only instance in which the Universe has hosted a technological species.

---

[3] Rather than $f_p$ and $n_p$ some authors focus on $\eta_E$, defined as the occurrence rate of Earth-size planets (1-2 Earth radii) in the habitable zones of solar-like stars. For example, one recent study based on Kepler mission transiting planets finds a value $\eta_E \sim 0.06$ (Silburt et al 2015). For the purposes of our order-of-magnitude estimates, which also include non-sunlike stars, we stay with the factors in the conventional Drake equation.

We can generalize these results for structural scales (of size $R_s$) within the Universe: galaxies, cluster of galaxies, and superclusters of galaxies (Fukugita et al 2004). Table 1 provides typical values for these entities, and lists the corresponding values of $f_{bt}$ for each of them, as well as for the entire Universe. For instance, for our own Galaxy $f_{bt}$ is $1.7 \times 10^{-11}$, meaning that another technical species has likely occurred in the history of the Milky Way if the probability of a technological species arising on *a given planet in a habitable zone* is greater than one in 60 billion. Figure 1 presents these results for given values of $f_{bt}$. The figure allows one to see the corresponding number of technological species that have ever arisen on various scales for various assumptions of the difficulty of biology and technology to evolve.

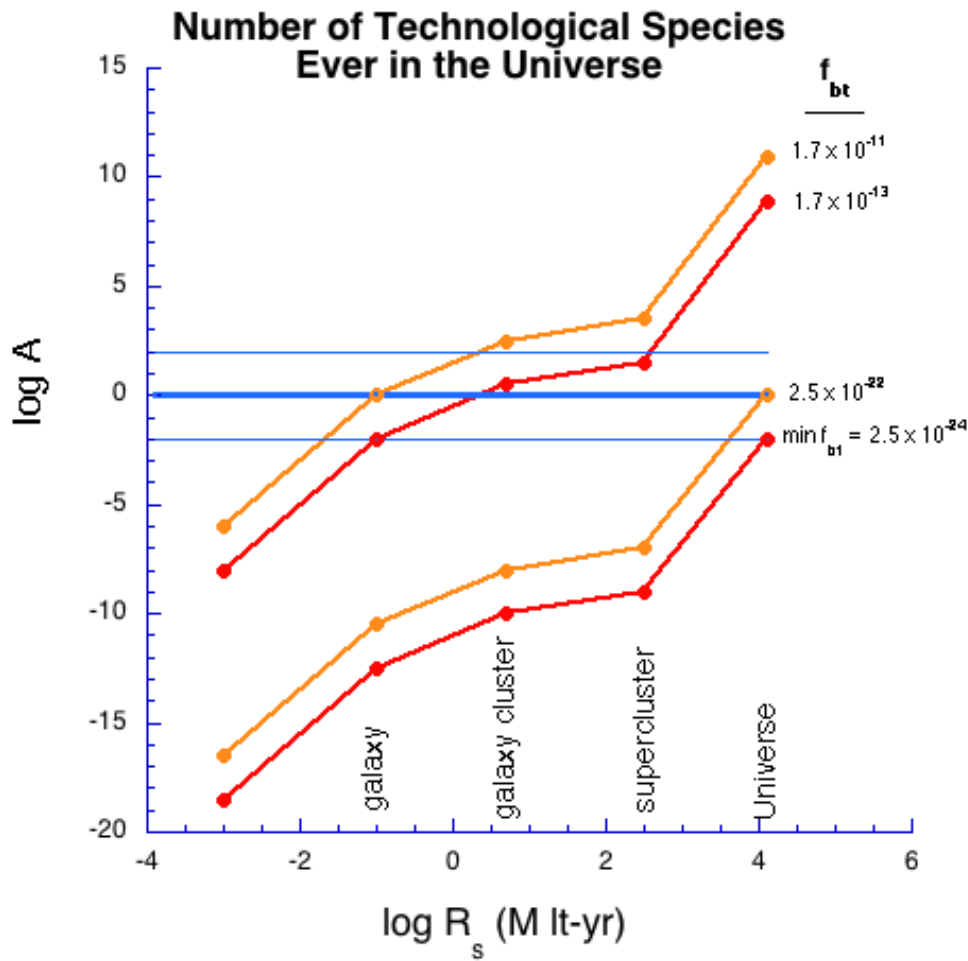

**Fig. 1.** The number *A* of technological species that have ever occurred over the history of the Universe, versus the scale size $R_s$ of different hierarchial structures in which their host stars are found: galaxies, clusters of galaxies, superclusters of galaxies, and the entire Universe. Reference values of *A* = 0.01, 1 and 100 are marked. The four curves represent different values of $f_{bt}$, the probability of a technological species arising on a given habitable zone planet. Only if $f_{bt}$ falls as low as $2.5 \times 10^{-24}$ to $2.5 \times 10^{-22}$ is it likely that no other technological species has ever arisen in the entire Universe.

**Discussion and Conclusions:** These results have wide implications. First, the long history of debate over extraterrestrial intelligence can be characterized as one of pessimists versus optimists relative to choices for the various factors in the Drake equation (Vakoch et al 2015) These debates have, however, been unconstrained in terms of how pessimistic one can be. With our approach we have, for the first time, provided a quantitative and empirically constrained limit on what it means to be pessimistic about the likelihood of another technological species ever having arisen in the history of the Universe. We have done so by segregating newly measured astrophysical factors from the fully unconstrained biotechnical ones, and by shifting the focus towards a question of "cosmic archaeology" and away from technological species lifetimes. Our constraint addresses an issue that is of particular scientific and philosophical consequence: the question "Have they ever existed?", rather than the usual narrower concern of the Drake equation, "Do they exist now?" Perhaps in the long term we should contemplate undertaking a field survey in cosmic archaeology, seeking possible evidence for such past technical species (Stevens et al 2015)

Finally, we note that sample sizes *A* of order 100 or 1000 have a particular importance in current discussions over efforts to create a sustainable, energy-intensive, high-technology civilization here on Earth (Frank & Sullivan 2014) Given the challenges human society faces from climate change, resource allocation, and loss of species diversity, it is not clear if the kind of long-term global culture we hope to build is even sustainable. However *A* >100 in our formulation implies that the evolution of technological species has occurred enough times that the ensemble of their histories (or trajectories in a suitably defined phase space) is statistically meaningful . In particular, for large enough values of $f_{bt}$ the average longevity <*L*> of such a sample of technological species does, in principle, exist.  Thus if $f_{bt}$ is such that *A* > 1000, it is reasonable to consider that different versions of humanity's current technological experiment in the alteration of our planetary system has occurred before. Note that many aspects of the feedback between an energy intensive, technological species and their host planet would depend solely on physical processes and constraints (atmospheric chemistry changes, hydrological cycle distruptions, etc; Frank & Sullivan 2014).  Thus modeling ensembles of such species to understand the broad classifications of their planetary feedback histories is theoretically well-grounded for large enough values of $f_{bt}$ . Such a project would be an attempt to understand both average properties like <L>, as well as what led some trajectories to collapse and others to long-term sustainability.

In conclusion we have shown that recent advances in astrobiology  and exoplanet studies mean that an empirically derived lower limit can now be placed on the probability that even one other technological species has ever evolved in the Universe.  This limit provides a framework for discussions of both life in its cosmic context and questions about trajectories of technological species relevant to our own issues of global sustainability.

**Acknowledgements.**  We thank Dan Watson and David Catling for comments on the text, and Matt McQuinn, Caleb Scharf and Gavin Schmidt for discussion.

# References

1. Impey, C., Lunine, J. & Funes, J. (eds.). *Frontiers of Astrobiology* (Cambridge Univ. Pr., 2012).

2. Sullivan, W.T. & Baross, J.A. (eds.). *Planets and Life: The Emerging Science of Astrobiology* (Cambridge Univ. Pr., 2007).

3. Vakoch, D.A. & Dowd, M.F. (eds.). *The Drake Equation: Estimating the Prevalence of Extraterrestrial Life through the Ages* (Cambridge Univ. Pr., 2015).

4. Cassan, A. et al. One or more bound planets per Milky Way star from microlensing observations. *Nature* **481**, 167-9 (2012).

5. Petigura, E.A., Howard, A.W. & Marcy, G.W. Prevalence of Earth-size planets orbiting sun-like stars. *Proc. Natl. Acad. Sci.* **110**, 19273-8 (2013).

6. Silburt, A, Gaidos, E, and Wu, Y,A Statistical Reconstruction of the Planet Population around Kepler Stars, *Astrophys. J.,* 799, 180, [2015]

7. Fukugita, M. & Peebles, P.J.E. The cosmic energy inventory. *Astrophys. J.* **616**, 643-68 (2004).

8. Stevens, A., Forgan, D. & O'Malley James, J. Observational signatures of self-destructive civilisations. ArXiv 1507.08530 (Jul 2015).

9. Frank, A. & Sullivan, W. Sustainability and the astrobiological perspective: Framing human futures in a planetary context. *Anthropocene* **5**, 32-41 (2014).

**Table 1   Limits on the Number of Technological Species**

**Occurring on Different Scale Lengths**

| Scale length | Size $R_s$ (M lt-yr) | No. Galaxies | $N_{ast}$ | $f_{bt}$ for $A$=1 |
|---|---|---|---|---|
| galaxy | 0.1 | 1 | 6 x $10^{10}$ | 1.7 x $10^{-11}$ |
| galaxy cluster | 5 | 300 | 2 x $10^{13}$ | 4 x $10^{-14}$ |
| supercluster | 300 | 3000 | 9 x $10^{15}$ | 2 x $10^{-15}$ |
| observable Universe | 13,700 | 7 x $10^{10}$ | 4 x $10^{21}$ | 2.5 x $10^{-22}$ |